# Effects of pulsed low frequency electromagnetic fields on water using photoluminescence spectroscopy: role of bubble/water interface

Philippe Vallée[a]

and

Jacques Lafait

*Laboratoire d'Optique des Solides (UMR CNRS 7601)*

Université Pierre et Marie Curie,

*Campus Boucicaut*

140, rue de Lourmel, 75015 Paris, France

Pascale Mentré

*UMR 8646 - MNHN*

*53 rue Buffon, 75005 Paris, France*

Marie-Odile Monod

*CEMAGREF*

*24, avenue des Landais - BP 50085*

*63172 Aubière Cedex, France*

Yolène Thomas

*Institut André Lwoff IFR89*

*7, rue Guy Moquet-BP8*

*94801 Villejuif Cedex, France*

[a] Author to whom correspondence should be addressed. Electronic mail: phvallee@los.jussieu.fr

# Abstract

The effects of a pulsed low frequency electromagnetic field were investigated on photoluminescence of well characterized water and prepared under controlled conditions (container, atmospheric, electromagnetic, and acoustic environments). When reference water samples were excited at 260 nm, two wide emission bands centered at 345 nm (3.6 eV) and 425 nm (2.9 eV) were observed. By contrast under 310 nm excitation, only one band appeared at 425 nm. Interestingly, electromagnetic treatment (EMT) induced, at both excitation wavelengths, a decrease (around 70%) in the 425 nm band relative photoluminescence intensity. However, no difference between reference and treated sample was observed in the 345 nm band. Other experiments, performed on outgassed samples (reference and treated), show that the emission bands (position, shape, intensity) under excitation at 260 nm and 310 nm were similar and close to the corresponding bands of the treated nonoutgassed samples. Similar effects were observed on photoluminescence excitation of water samples. Two excitation bands monitored at 425 nm were observed at 272 nm and 330 nm. After EMT and/or outgassing, a decrease (> 60%) was observed in the intensity of these two bands. Altogether, these results indicate that electromagnetic treatment and/or outgassing decrease in a similar fashion the photoluminescence intensity in water samples. They also suggest that this effect is most likely indirectly attributed to the presence of gas bubbles in water. The possible role of hydrated ionic shell around the bubbles in the observed extraluminescence is discussed.

# Introduction

The effect of static magnetic and electromagnetic fields (EMF) on water or on the behavior of aqueous solutions and suspensions has increasingly drawn the attention of investigators in recent years[1-5]. For instance, it has been shown that water exhibits changes in physicochemical properties in response to variations of field intensity and/or frequency[4,5]. The idea of a change in the structure of water itself, as a result of magnetic exposure has been criticized due to the low energy from the used field intensities in comparison with the one from thermal agitation. The situation is complicated by the fact that most of the effects studied depends on water composition and/or environmental conditions. Although the precise mechanisms of the EMF effect has not been yet fully elucidated, some theoretical and experimental considerations have suggested that EMF act *via* a colloid/water interface[6-8]. In this regard, Colic and Morse[6] working with oscillating radio-frequencies EMF on colloids in water proposed that the electromagnetic treatment (EMT) effect results from the perturbation of the gas/liquid interface and that outgassing removes all the observed effects which persist for hours and even days (Zeta potential, turbidity, …)[9]. This interface has been shown to play a key role in emulsion stability, cavitation, and biological reactivity[10]. Fesenko and Gluvstein[11] treating triply distilled deionized water with microwaves suggested that the EMT might act on gas dissolved in water. Eberlein[12] proposed a quantum vacuum radiation explanation of the effect based on the influence of an oscillating EMF on the gas/water interface. According to her theory, a moving interface between media of different polarizability (water and bubbles) is a possible two-photon state source under the excitation by the electromagnetic field[13,14]. Other studies[15,16] showed that the effect of magnetic field on fluorescent probes is affected by the solvent nature and their environment. For instance, Higashitani *et al.*[15] followed the fluorescence emission intensity of probes dissolved in water-ethanol mixtures exposed to a static magnetic field. The authors[15] suggested that the static magnetic treatment altered the orientation of the water molecules around the alkyl chain of the fluorescent probes and also the conformation of the water molecules, ions, hydrated ions adsorbed on the particle surface. Chowdhury *et al.*[16] showed that the magnetic field modulated exciplex luminescence in liquids. The exciplex luminescence depends on the liquid structure and the environment, such as viscosity, dielectric constant of the medium and presence of other molecules in the neighborhood of the exciplex.

In a previous work[17], we have shown that the optical properties of highly purified water can be altered by the photoluminescence of traces impurities arising from container/content interactions. Therefore, in order to investigate the effects of pulsed low frequency of electromagnetic treatment on water photoluminescence (PL), we made some attempts to develop new experimental conditions, in particular to minimize the release of compounds from the containers. For instance, glassware was made of pure fused optical silica and water was purified under controlled environmental conditions (atmospheric, electromagnetic and acoustic environments). In the experiments reported here, we show that electromagnetic treatment induces a modification in the relative photoluminescence intensity of the water samples. We will discuss the possible mechanisms at the origin of this effect.

## 1. Materials and methods

### 1.1 Reagents

*Water*

The water used in the experiments described below was freshly prepared with Paris tap water (pressure 6 bar) after two purification steps using: (1) an inverse osmosis apparatus (Rios III, Millipore) that eliminates the molecules between 0.2 and 1 nm, 97% of the ionic substances as well as 99% of the organic compounds; (2) a final polishing apparatus (Simplicity, Millipore) equipped with an ultrafilter (Pyrogard D, size cutoff of 13 000 Da, Millipore) that eliminates the remaining organic compounds ≥ 5-10 nm. At the end of the procedure, water (purified water) was apyrogen according to the certificate of the manufacturer and the resistivity was 18.2 MΩ.cm at 25°C.

*Ethanol*

Spectroscopy grade ethanol (purity 0.999, UVASOL) purchased from Merck was used to clean the glassware. Moreover in order to minimize the fluctuations the residual concentration of organic compounds (≈ 0.05 ppm), a small amount of the same ethanol was added (≈ 0.95 ppm) during water preparation. According to the manufacturer, ultraviolet transmittance is superior or equal to 98% and 100% at 270 nm and 310 nm respectively. Importantly, at 254 nm and 365 nm wavelengths respectively, the fluorescence intensity of ethanol is inferior or equal to that of a quinine solution of 2 ppb and 1 ppb (or any other fluorescent compounds).

*Glassware*

All glassware were made of pure fused optical silica (Suprasil, Heraeus, and Hellma) in order to minimize container/content interaction[17,18]. Optical cells (v: 3.5 ml, section: 1cm$^2$), closed also with a pure fused silica cap, were purchased from Hellma (QS 111). Glassware was used specifically and only for the study. All glassware were thoroughly washed with ethanol to remove any surface active material and then put in a clean oven at 80°C for 15 min. Just before use, the optical cells and Erlenmeyer 250 ml flask were rinsed with purified water. A specially designed glassware apparatus made of two compartments, a round beaker connected to an optical cell was used to outgassed water.

*Glove box*

The glove box was purchased from Jacomex (large-scale model: B003). In order to reconstitute atmospheric air, argon (N56, Air Liquide), carbon dioxide (N48, Air Liquide), oxygen, and nitrogen (Air N57 Pol, Air Liquide) were filtered through a 0.2 μm filter (Teflon Calyx Capsule,

Osmonics) and then introduced into the glove box. A 3 mbar above the atmospheric pressure was maintained in order to avoid outside contamination. Relative humidity (60 ± 2%) and carbon dioxide concentration (310 ± 20 ppm) were measured using a GM70 hand-held purchased from Vaisala (calibrated probe, humidity: HMP75 and carbon dioxide: GMP70).

*Insulated cages and storage boxes*

Two cylinder insulated cages (diameter 700 mm, 500 mm high) were designed and custom built. For thermal and acoustic insulation, we used multi-layers made of various polymers (EPDM, density: 8kg/m$^3$, 20 mm thick; melamine resin, density: 11kg/m$^3$, 20 mm thick and bitumen mass, density 10kg/m$^3$, 3 mm thick, Illbruck company). For magnetic insulation, we used a mu-metal double foil, 1mm thick for each foil (Mumetal, Meca-Magnetic). According to the manufacturer, magnetic field intensity (0.1 mT, ac, 50 Hz) is diminished by a factor of 850 ($\approx$ 58 dB). For grounding electric insulation ("Faraday cage"), copper foil (2 mm thick) was used.

Two storage boxes were specially designed, composed of a shielding closed cylinder made of mu-metal (diameter 80 mm, 105mm high, 2 mm thick) and placed in a Thermos® box.

*1.2 Water treatments*

*Electromagnetic treatment*

Optical cells containing purified water were placed in one of the two insulated cages and exposed for 6 h to pulsed (duration 30 s) electromagnetic fields generated by a solenoid coil (diameter 50 mm, 80 mm high, copper wire, 4367 turns/m, self inductance L = 3 mH, ohmic resistance 3 Ω). The signal produced by a function generator (Agilent 33120A), consisted of 3 successive sinusoidal signals in the frequency range from 10 to 500 Hz. Optical cells were placed in a vertical position in the center of the coil supplied by a 250 mA rms current. The calculated r.m.s. magnetic field density at the center of the coil was $\approx$ 1 mT and according to Faraday law, the maximum induced electric field was 4.1 mV/m. Control untreated (reference) optical cells were placed in the other insulated cage under the same conditions as exposed optical cells. After the period of EMT, optical cells [reference (R)/treated (T)] were transferred and kept into separate storage boxes. During the electromagnetic treatment, the temperature had been monitored using a chromel/alumel thermocouple ($\Delta\vartheta$ was less than 0.01°C).

*Outgassing process*

To outgass purified water, the technique described by Pashley and co-workers[19,20] which gives a degassing efficiency of $\approx$ 97% was used with a slight modification. A zeolithe pump instead of a water jet pump aspirator was used to outgass water (20 min, 12 torr, pressure measured by a manometer C9555, Comark). At the end of pumping, outgassed water was transferred from the Suprasil round beaker to the optical cells [outgassed reference water (ROG) and outgassed treated water (TOG)]. Of note, under our experimental conditions, the glassware apparatus could not be entirely filled up to the top.

*1.3 Measurements*

*Physicochemical water analysis*

The different measurements, using a multiparametric system (multimeter Consort C835, Fisher Scientific ) were always done, in the glove box, in the same order: conductivity/temperature, oxygen concentration, pH, and oxidation-reduction potential (ORP). Data values from the multimeter were recorded on a computer. For conductivity measurements, the apparatus was equipped with a specific electrode of cell constant K=0.1 cm$^{-1}$ for low ionic concentration solutions with automatic temperature correction (Pt1000) and calibrated with two ionic strength solutions (84 and 1413 µS.cm$^{-1}$ at 25 °C, Hanna Instruments). pH measurements were done using a combined electrode (LL-Aquatrode, Metrohm) adapted for pure water. The electrode was calibrated using a test kit buffer (Scott-Geräte). For ORP measurements, we used a combined electrode with platinum ring (N90417, Fisher Scientific) and for oxygen measurements, we used a combined electrode oxygen/temperature (N98338, Fisher Scientific). The ORP probe was calibrated using a redox solution of 470 mV (Pt-Ag/AgCl at 25°C, Scott-Geräte) and the oxygen probe calibrated with 100% of oxygen content in air. Analysis of Total Organic Compounds (TOC) was measured by an A10 Monitor (Anatel, Millipore). The accuracy was 15% in the range of 2-1000 ppb.

*Ambient magnetic fields (geomagnetic and environmental) analysis*

The dc and ac electric as well as magnetic fields were measured inside of the insulated cages with a low field axial probe Mag B, positioned at 60°, connected to a Mag-01H (Bartington Instruments Ltd) for the dc measurements. For ac analysis, measurements were done along three axes over a frequency range of 5 Hz to 400 kHz with the probe of apparatus Esm-100 (Maschek) placed close to the exposure area. In the absence of current, the fields in the solenoid coil were: static magnetic field 9.9 µT, ac magnetic field (in the range 5 Hz to 400 kHz) 16 nT, ac residual electric field (mainly 50 Hz, isotropic) 6.3 V/m. Of note, the applied pulsed magnetic field (1 mT), was far above the environmental magnetic fields both ac (16 nT) and dc (less 10 µT). The maximum induced electric field produced by the applied electromagnetic field, 4.1 mV/m, is very weak compared to the environmental electric field (few V/m).

*1.4.    Photoluminescence analysis*

Emitted light from the sample was recorded at right angle of the excitation beam, on a Fluorolog3-2-Triax spectrofluorometer (Jobin Yvon). The light source was a xenon lamp (450 W) operating in a wavelength range of 240 to 850 nm. The excitation wavelength was selected by a double monochromator (gratings 330 blaze, 1200 grooves/mm, dispersion: 2.1 nm/mm). A Triax 320 spectrometer was used to analyze the emission spectra. It is equipped with a monochromator (grating 750nm blaze, 600 grooves/mm) and a photomultiplier Hamamatsu R928. The spectra were corrected from the source fluctuations. Two excitation wavelengths (260 and 310 nm) were chosen in order to compare our results with Lobyshev's[21]. We will see further that they are close to the optimal excitation wavelengths. For the 260 nm wavelength (40 mW power) and 310 nm (190 mW power), the explored emission wavelength ranged from 270 nm to 500 nm and 325 nm to 600 nm, respectively. In the excitation spectra experiments, the excitation spectra ranged from 240 to 330 nm and 240 to 410 nm, respectively, according to the emission wavelength monitored at 345 nm and 425 nm, respectively. For both excitation and emission spectra, resolution of the PL spectrum was 5 nm with an increment of 1 nm and an acquisition time of 1s per increment. In a preliminary work, we noticed, in agreement with Lobyshev's observations[21], an increase in the PL intensity of our reference samples as a function of time. We also observed that EMT samples presented over time the same behavior but with a slower increase in the PL intensity. In order to maximize the differential effect, we decided to present here the experimental results obtained three weeks after sample preparation and treatment. Of note, during the period, each sample was kept in individual storage box.

## *Results*

We standardized on a common experimental procedure and used the same exposure device in all the tests we performed (Fig. 1). In brief, all glassware was made of purified fused silica; water was purified and samples prepared inside a glove box under a controlled atmosphere. In order to have a good homogeneity among different samples, water coming from the "Simplicity apparatus" was stored for 25 min in a closed Erlenmeyer flask. Then, water was quickly poured into the optical cells and filled to the top. Optical cells were closed and sealed with a silica ground cap. During this period, an aliquot sample was taken and analyzed for physicochemical measurements. All control (reference) and treated water samples were made of purified water having the following characteristics: resistivity 4.0 MΩ.cm at 25°C, TOC content: ≈ 1 ppm (mainly ethanol), pH: 5.9, redox potential: 280 mV, oxygen content: 2 ppm.

We first investigated the effect of a 6 h pulsed low frequency electromagnetic treatment on water using photoluminescence spectroscopy. Of note, the intensity of the Raman peak for each couple (reference and treated samples) remains constant. In order to compare results between experiments, each spectrum was normalized to the maximal values of the Raman. Figure 2 depicted the emission spectra of water samples excited at a wavelength of 260 nm (panel A) and 310 nm (panel B). Wide structureless bands centered at 345 nm (3.6 eV) and 425 nm (2.90 eV) and only one band at 425 nm were observed for 260 and 310 nm excitation wavelengths respectively. Of note, in the reference samples (R), the amplitude of the 425 nm emission band in reference samples decreases by about 50% when excited at 310 nm as compared to 260 nm. The full width at half maximum (FWHM) of the 425 nm band (excited at both wavelengths) is ≈ 55 nm. Interestingly, after electromagnetic treatment (T), we noted, at both excitation wavelengths, a striking decrease (around 70%) of the 425 nm band. By contrast, no variation between R and T samples was seen in the 345 nm band. Other experiments (Fig. 3), performed with outgassed samples (ROG, TOG), show that the emission bands (position, shape, intensity) under excitation at 260 nm (panel A) and 310 nm (panel B) were very similar and close to the corresponding bands of the treated nonoutgassed samples (Fig. 2). Altogether, these results indicate that electromagnetic treatment or outgassing decrease in a similar manner the photoluminescence intensity of water.

The photoluminescence excitation (PLE) of water samples was next studied. As before, each spectrum was normalized to the maximal values of the excitation band corresponding to the Raman peak. Figure 4 shows the excitation spectra of samples R and T (panel A) and ROG and TOG (panel B) monitored at 345 nm. One wide band of low intensity was observed around 280 nm. The pattern of the excitation spectra was roughly similar for each couple (R *vs* T) and (ROG *vs* TOG). This latter result was not surprising if one considers the similarity of the emission bands around 345 nm (excited at 260 nm) for each couple. By contrast and as illustrated in Fig. 5, when monitored at 425 nm, two excitation bands peaking at 272 and 330 nm were present. Of note, the PL 425 nm band did not present mirror symmetry with the excitation bands at 272 and 330 nm. Interestingly, after EMT or outgassing, a 65% relative decrease was observed in the PLE intensity of these two bands. Of note, the larger stokes shift around 150 nm (13 235 cm$^{-1}$) may suggest a reorganization of the whole electronically excited molecules before the emission takes place[22]. These results again indicate that electromagnetic treatment or outgassing decrease in a similar fashion the PLE intensity of water.

## *Discussion*

The intensity of the 425 nm emission band observed in reference water samples could not be merely attributed to impurities alone coming either from the container/content interaction or simply present in water since: (i) all glassware were made of pure fused optical silica; (ii) the 1 ppm spectroscopy grade ethanol present in water did not interfere with the ultraviolet transmittance at 310 nm wavelength (see 1.1). (iii) the water purification process (inverse osmosis plus Simplicity apparatus equipped with a Pyroguard ultrafilter) guarantees the lowest amount of impurities. In addition, the water used in the experiments was purified under controlled environmental conditions and always displays the same physicochemical characteristics (Fig. 1). Interestingly, our data also indicate that outgassing decreased (by about 70%) the photoluminescence intensity in the reference samples. The latter data also pointed out the role of gas bubbles in the observed effect. We would emphasize, however, that it is difficult to formally rule out at present , the possibility that trace amount of residual impurities in our samples may have contributed to the 5-9% relative emission intensity, observed after either EMT or outgassing (Fig. 3).

The precise origin of the 70% "extra-photoluminescence" intensity remains a puzzling and open question. Although this point cannot be resolved at the present time, at least two possibilities can be readily envisioned. For example, it is conceivable that the 425 nm emission spectrum of the reference samples might be due to intrinsic photoluminescence of water, as proposed by Lobyshev *et al.*[21]. An alternative view is to assign the origin of the extra-photoluminescence intensity to trace amount of impurities adsorbed at the bubble/water interface. We favor the latter interpretation since, outgassing water decreases the photoluminescence intensity (by about 70%) in the water samples. In addition, we have recently identified using dynamic light scattering, the presence of gas nanobubbles (around 300 nm) in non-outgassed water samples, that partially vanished under the action of EMT (Ph. Vallée *et al.*, Langmuir 2005, in press). Furthermore, our experimental conditions to prepare purified water are very close to a compression-expansion cycle: relative high pressure (6 bar) at the entrance of the inverse osmosis apparatus and low pressure (1 bar) at the exit of Simplicity apparatus. A state of gas supersaturation is therefore reached by the water, which may lead to the formation of gas bubbles in our samples, stable over a long period of time, since our cells are filled to the brim and hermetically sealed. In support of this notion, Bunkin and co-workers[23-27] working on cavitation of gas bubbles in liquids, suggested that gas nanobubbles occur and are stabilized by a shell of negative ions present in the water, even when ion concentration is as low as 0.1 ppm. Indeed, such concentration can be found in our water samples, due to (i) water autodissociation ($OH^-$ ions)[28]; (ii) carbon dioxide gas present in water, indeed, resistivity (4.0 MΩ.cm) and pH (5.9) measurements before closing the optical cells correspond to a $HCO_3^-$ content of ≈ 0.1 ppm; (iii) a release of trace silica ions from the optical cell during the storage.

An interesting finding that became apparent during these studies, is that electromagnetic treatment of water yields nearly similar results as outgassing. In experiments close to ours, on an aqueous electrolyte solution, Beruto et al.[29] showed that low frequency electromagnetic fields (250 µT, square pulsed fields at 75 Hz), have the same effect (vaporization of carbon dioxide) as outgassing. As a consequence, ionic concentration decreases in the solution and tends to neutralize the interfacial area. Similarly, our pulsed low frequency electromagnetic treatment could act on the bubble/water interface leading to destabilization of bubbles, in particular by disturbing the ionic balance between the shell of adsorbed negative ions and counter ions. Others authors (Gamayunov[30] and Lipus[31]) envision the action of EMF directly on the ionic doublelayer. Along this line, Gamayunov et al.[30] show that the Lorentz force causes a local deformation of the electrical double-layer particles or air bubbles of various sizes resulting into the coalescence of the gas bubbles, when they are carried by the liquid in a static magnetic field (around 75 mT). Furthermore, Lipus et al.[31] also explained the effect of static magnetic treatment (intensity range 0.05 – 1 T) on the dispersion of particles or bubbles in water by the Lorentz force which induces the shift of counterions from the Gouy-Chapman into the Stern layer.

Let us now consider different hypotheses regarding the relationships between the decrease in PL intensity and the disappearance of gas bubbles. Most of them are related to the electric charge density of ionic compounds present at the bubble/water interface which has been shown to play a crucial role in colloids, beginning with Hofmeister works[32,33]. Indeed, the molecular interactions occurring in the ionic shell present at the bubble surface could produce collective electronic effects being thus possibly at the origin of the observed PL. This effect has been also observed in solid compounds. Su and Guo[34] explained the PL intensity changes observed in hydrogenated amorphous silicon oxide powder by the modification of the electronic density of states of silicon due to defects and variations of the local environment. By the same token, the analysis of various published data[35,36] shows that a photoluminescence band around 410 nm excited at 337 nm is ascribable to the presence of defects (like Si ions) in silica samples. Our results concerning the emission band of water samples around 425 nm when excited at 310 nm (optimally at 330 nm) may relate to previous observations that ascribe a key role of silica ions adsorbed on bubbles. Even if there is a low release of silica ions after a long-term storage in silica cells, these ions can be adsorbed preferentially on the bubble/water interface, thus increasing the electric charge density that might contribute to photoluminescence. Ionized ethanol molecules preferentially adsorbed at the bubble/water interface may act in the same way by inducing a greater charge density at the bubble surface.

As discussed above, it is unlikely that the extra-photoluminescence intensity observed may be due to the intrinsic photoluminescence of the ethanol molecule alone. Nevertheless, due to the ethanol behavior as proton trap[37-39], a proton transfer between residual organic compounds and hydrated OH groups of the ethanol, both at the bubble/water interface may occur and contribute to the observed extra-photoluminescence intensity in reference samples. The proton transfer rate can vary with the solvent composition due to the variation of its dielectric properties. Along this line, Chowdhury[16] showed that variations of the luminescence of exciplex depends on the structure of the liquid and on the dielectric constant of the solvent. Moreover, Lepoint-Mullie *et al.*[40] following the Gouy-Chapman model, infer that the dielectric constant of water is decreasing when approaching the surface of bubbles.

## *4. Conclusions*

Several points emerge from this study:
(i)  A thorough process of water sample preparation and characterization to measure the action of pulsed low frequency electromagnetic fields on photoluminescence of purified water was developed.

(ii) Under 260 nm excitation wavelength, two wide structure-less photoluminescence bands at 345 nm and at 425 nm were observed, while only one band at 425 nm appeared under 310 nm excitation.

(iii) After electromagnetic treatment and/or outgassing, the 425 nm band emission intensity decreases sharply (around 70%).

(iv) When the PLE spectra was monitored at 425 nm, two excitation bands peaking at 272 nm and 330 nm were observed. After EMT and/or outgassing, a decrease (around 65%) was observed in the PLE intensity.

(v) Among these different models discussed above, we favor, at the present time, the possibility that the 70% extra-photoluminescence intensity observed may be due to the electronic charge density induced by the concentration of hydrated ionic compounds around gas bubbles, on which electromagnetic treatment acts. Clearly, further studies will be necessary to directly address these issues.

These results may open interesting perspectives in biology, notably via the role of the ionic double layer[41,42].

## *Acknowledgements*

This work was part of a Ph.D. thesis prepared by Ph. Vallée and was supported by a grant from the Odier Fondation.

The assistance of R. Nectoux, Ph. Camps, and G. Vuye in the design and realization of the electromagnetic device was highly appreciated. We also thank B. Démarets, S. Chenot, and M. Lempereur for technical assistance. We are indebted to C. Barthou for help in photoluminescence measurements and procedures. We are especially grateful to R. Strasser, B. Cabane, B. Guillot, I. Tatischeff and M. Odier for helpful advices and fruitful discussions.

**Figure captions:**

**Figure 1**. **Schematic drawing of the experimental procedures.**

Water was purified and samples prepared inside a glove box under a controlled atmosphere. Water samples (R, "reference sample"; T, "pulsed low frequency EMF treated sample"; ROG, "reference sample after outgassing"; TOG, "pulsed low frequency EMF treated sample after outgassing") have the following physicochemical characteristics: resistivity: 4.0 MΩ.cm at 25 °C, pH: 5.9, redox potential: 280 mV, oxygen concentration: 2 ppm.

**Figure 2: Photoluminescence (PL) emission spectra of nonoutgassed water samples: R, reference; T, after electromagnetic treatment.**

Each spectrum was normalized to the maximal values of the Raman peak. **Panel A:** excitation wavelength: 260 nm. **Panel B:** excitation wavelength: 310 nm.

**Figure 3: Photoluminescence (PL) emission spectra of outgassed reference (ROG) and outgassed treated (TOG) water samples**

Each spectrum was normalized to the maximal values of the Raman peak. **Panel A:** excitation wavelength: 260 nm. **Panel B:** excitation wavelength: 310 nm.

**Figure 4: Photoluminescence excitation (PLE) spectra, monitored at 345 nm**

"Raman peak": excitation band (observed at 310 nm) corresponding to the Raman peak centered at 345 nm. Each spectrum was normalized to the maximal values of the excitation band. **Panel A:** reference (R) and treated (T) water samples. **Panel B:** outgassed reference (ROG) and outgassed treated (TOG) water samples.

**Figure 5: Photoluminescence excitation (PLE) spectra, monitored at 425 nm**

"Raman peak": excitation band (observed at 372 nm) corresponding to the Raman peak centered at 425 nm. Each spectrum was normalized to the maximal values of the excitation band. **Panel A:** reference (R) and treated (T) water samples. **Panel B:** outgassed reference (ROG) and outgassed treated (TOG) water samples.

*Figures:*

**Figure 1:**

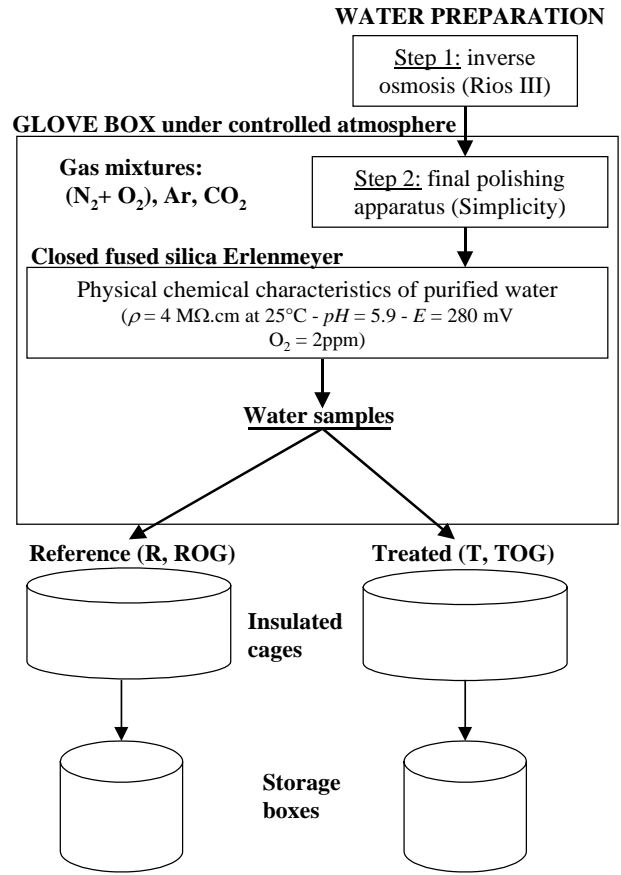

**Figure 2.**

Panel A                                    Panel B

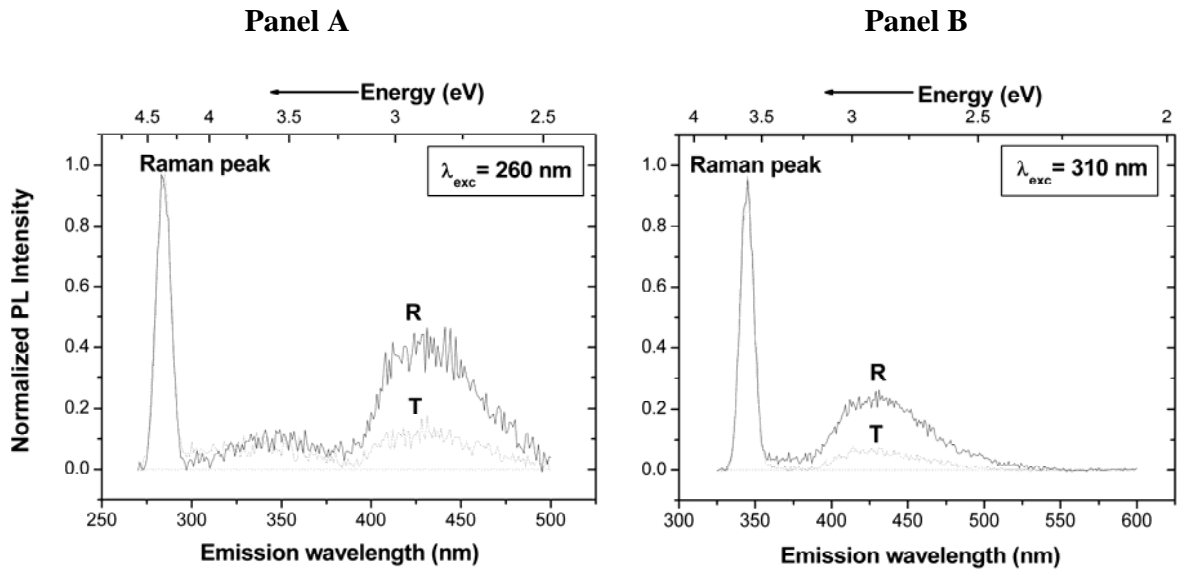

**Figure 3.**

Panel A                                    Panel B

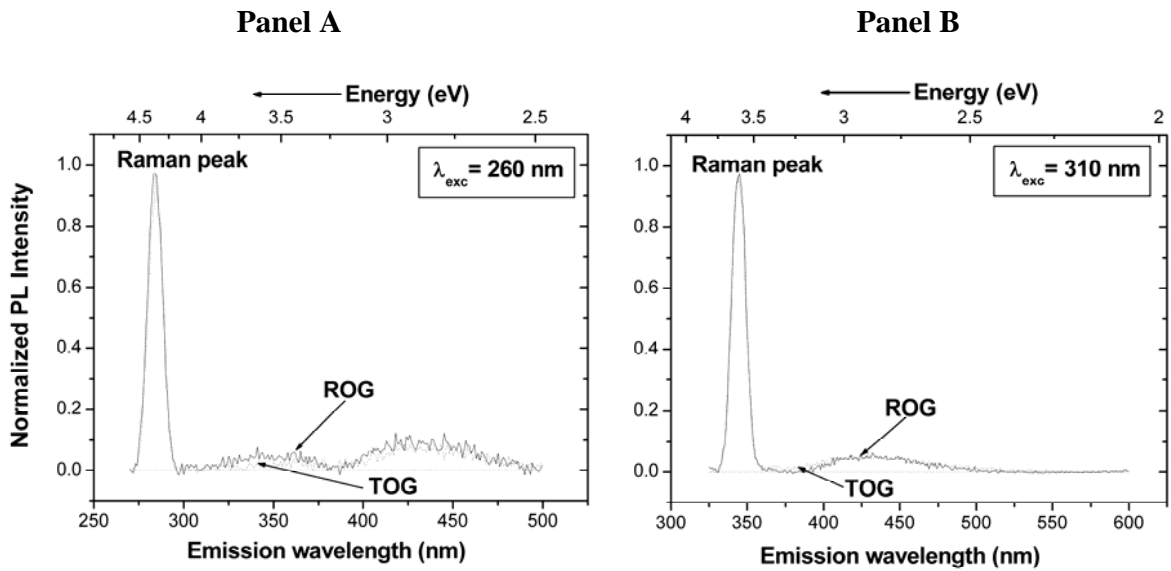

**Figure 4.**

Panel A                                           Panel B

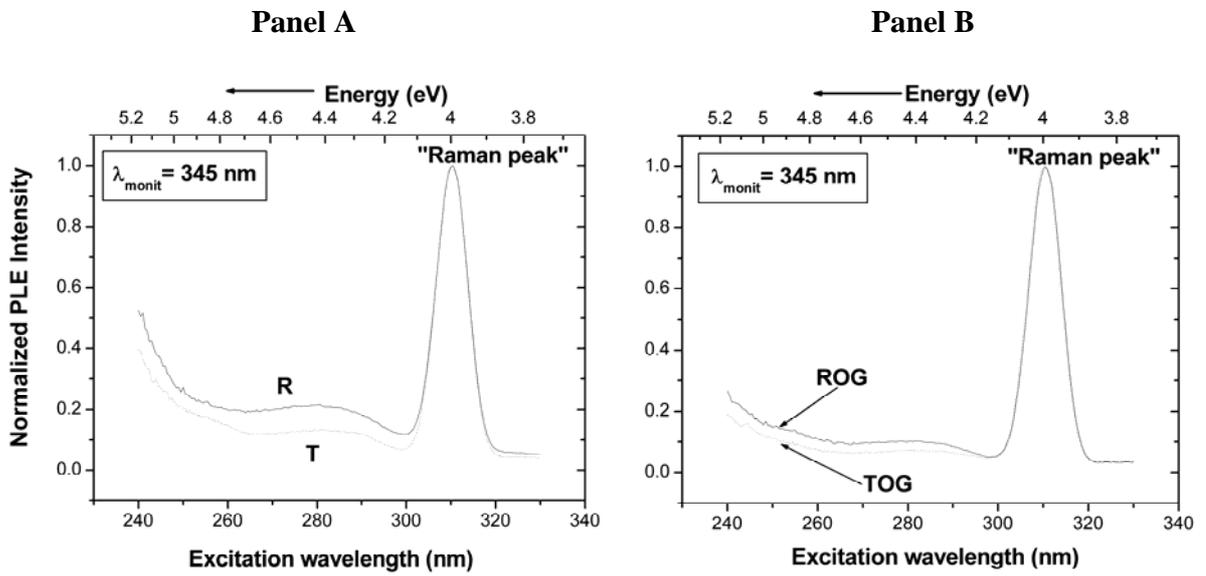

**Figure 5.**

Panel A                                           Panel B

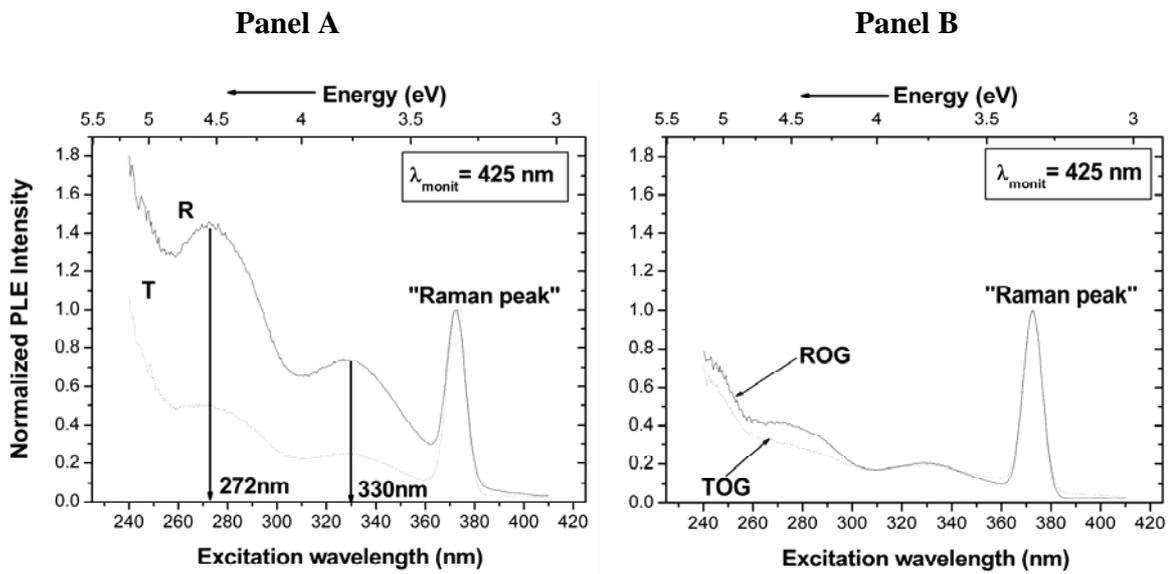